%% file: egpaper_for_review.tex
\documentclass[10pt,twocolumn,letterpaper]{article}

\usepackage{iccv}
\usepackage{times}
\usepackage{epsfig}
\usepackage{graphicx}
\usepackage{amsmath}
\usepackage{amssymb}
\usepackage{arydshln}
\usepackage[accsupp]{axessibility}

\usepackage[pagebackref=true,breaklinks=true,letterpaper=true,colorlinks,bookmarks=false]{hyperref}

\iccvfinalcopy 

\def\iccvPaperID{3926} 

\ificcvfinal\fi

\newcommand {\bt}[1]{\bf{#1.}\normalfont}

\newcommand{\net}{\mathcal{N}}
\newcommand{\disc}{\mathcal{D}}

\newcommand{\pix}{\mathbf{x}\normalfont}

\newcommand{\lum}{Y}
\newcommand{\lumldr}{Y_L}
\newcommand{\loglum}{\lum_{c}}

\newcommand{\loss}{L}
\newcommand{\lumfac}{\lambda}
\begin{document}

\title{Unpaired Learning for High Dynamic Range Image Tone Mapping}

\author{Yael Vinker
\qquad
Inbar Huberman-Spiegelglas
\qquad
Raanan Fattal\\
{\tt\small \{yael.vinker, inbar.huberman1, raanan.fattal\}@mail.huji.ac.il}\\
School of Computer Science and Engineering \\ The Hebrew University of Jerusalem, Israel\\

}

\maketitle
\ificcvfinal\thispagestyle{empty}\fi

\begin{abstract}

High dynamic range (HDR) photography is becoming increasingly popular and available by DSLR and mobile-phone cameras. While deep neural networks (DNN) have greatly impacted other domains of image manipulation, their use for HDR tone-mapping is limited due to the lack of a definite notion of ground-truth solution, which is needed for producing training data. 

In this paper we describe a new tone-mapping approach guided by the distinct goal of producing low dynamic range (LDR) renditions that best reproduce the visual characteristics of native LDR images. This goal enables the use of an \emph{unpaired} adversarial training based on unrelated sets of HDR and LDR images, both of which are widely available and easy to acquire.

In order to achieve an effective training under this minimal requirements, we introduce the following new steps and components: (i) a range-normalizing pre-process which estimates and applies a different level of curve-based compression, (ii) a loss that preserves the input content while allowing the network to achieve its goal, and (iii) the use of a more concise discriminator network, designed to promote the reproduction of low-level attributes native LDR possess.
 
Evaluation of the resulting network demonstrates its ability to produce photo-realistic artifact-free tone-mapped images, and state-of-the-art performance on different image fidelity indices and visual distances.
\end{abstract}


\input {Introduction}

\input {PrevWork}

\input {Method}

\input {Results}

\input {Conclusions}
{\small
\bibliographystyle{ieee_fullname}
\bibliography{egbib}
}
\end{document}

%% file: Introduction.tex
\section{Introduction}

High dynamic range (HDR) photography gained a considerable popularity in the last decades among both professional and non-professional photographers. HDR capabilities are available in many DSLR cameras as well as in mainstream mobile smartphones, capable of providing 10- and 12-bits of color depth. Printing or displaying these images on conventional low dynamic range (LDR) devices require a tone-mapping step for reducing their dynamic range. The latter revealed itself as a non-trivial task and drew a considerable research effort.

Unlike many image processing and restoration tasks, there is no ground-truth ``solution" for tone-mapping HDR images, and the different methods developed over the years aim for different goals. Early approaches use global tone-reproduction curves (TRC) that make a better use of the output dynamic range than linear bracketing~\cite{Drago2003,Larson97,Tumblin93}. These curves avoid over- and under-exposed pixels, but their contractive nature leads to a severe reduction in local contrasts. 

Consequently, more modern approaches focus their goal on preserving, or enhancing, the local contrasts using detail separation and enhancement techniques~\cite{Durand02,Farbman08,Fattal02,Paris11,Pattanaik98,Reinhard02,Tumblin99}. These methods produce highly-detailed images, but applying high levels of compression remains a challenge in terms of avoiding edge-related artifacts or achieving balanced levels of contrast and an overall photo-realistic appearance.

\begin{figure}
\centering
  \includegraphics[width=1\linewidth]{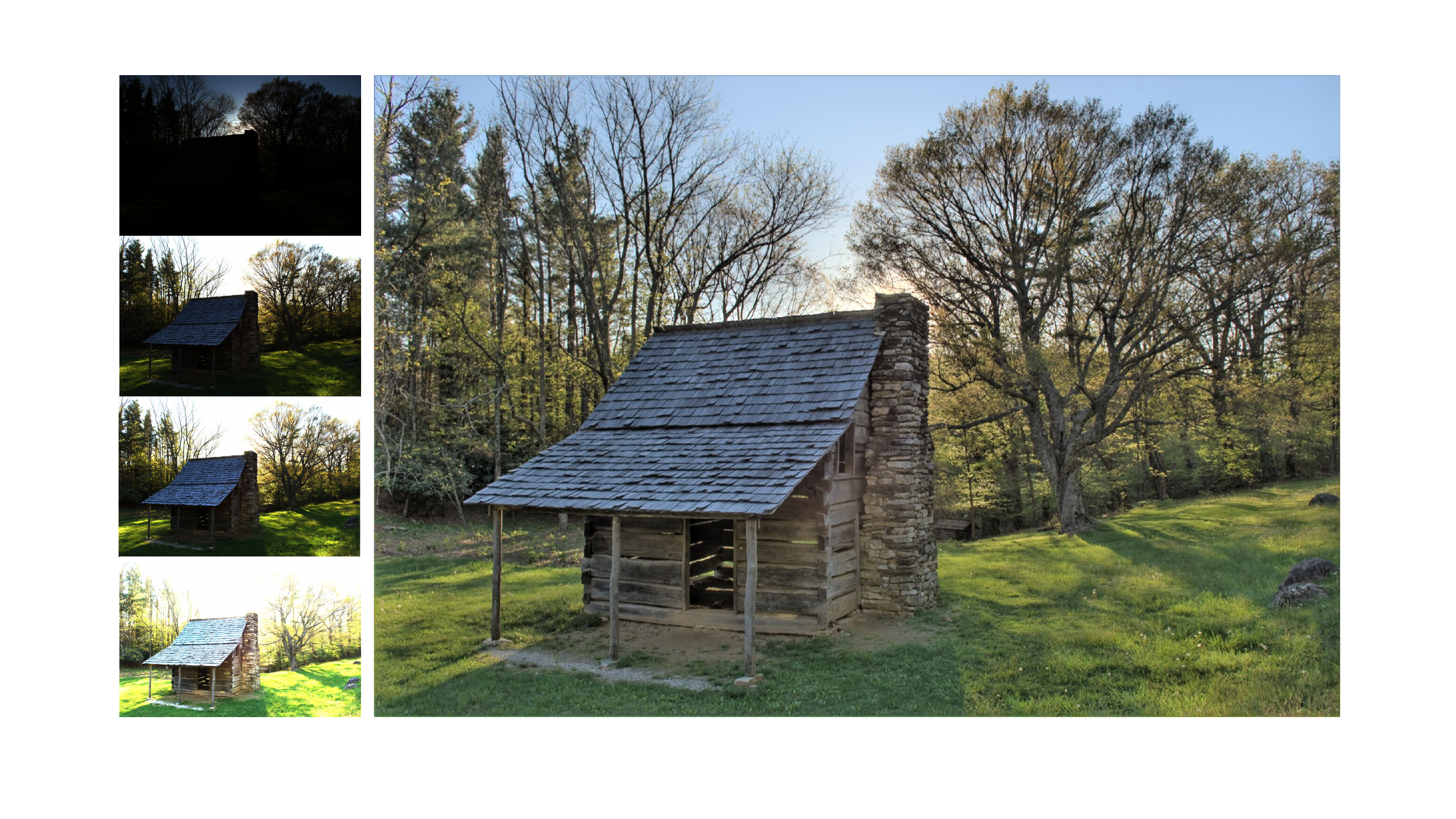}
  \caption{Tone-mapped HDR images produced by our method. The four exposures on the left of each image portray the very high dynamic range in the original scenes. \vspace{-0.1in}}
  \label{fig:teaset}
\end{figure}


Deep neural networks (DNNs) have greatly impacted various image processing tasks, such as super resolution~\cite{Johnson16}, and deblurring~\cite{Shen18}, by using large training sets containing ground-truth examples. In the absence of ground-truth data, the prevalent approach in HDR tone-mapping is using existing tone-mapping operators (TMOs) to produce a panel of image labels, and narrowing it down using an image quality index to obtain the final training examples~\cite{Patel2018, Rana20}.

The trained network is expected to reproduce the best \emph{available} result on each image, but not surpass the performance of its underlining TMO algorithms. Moreover, the quality indices used reward for fulfilling a small number of regularities and hence they bias the training towards overfitting these attributes. Some approaches incorporate manual supervision, but they still rely on the indices in their final selection~\cite{Zhang19}, or make a subjective decision by picking a single annotator~\cite{montulet2019deep}. 

In this paper we describe a new DNN-based TMO which is trained to produce images that bear the visual characteristics found in native LDR images. By formulating this distinct goal as an adversarial training, we replace the need for obtaining paired training examples with the plenitude of available high-quality  LDR images. 

Unsupervised adversarial training is a rather delicate process, prone to various instabilities. In order to achieve an effective and successful training, our method uses the following several new steps and components: (i) Before feeding the input HDR image to the network, we map its luminance through a range compression curve that reduces its variance and fixes its range. In order to train and apply our method on images from arbitrary sources, we use an adaptive compression level which we estimate on the basis of each input image.

Moreover, (ii) we incorporate a structure-preservation loss that penalizes for changes beyond local adjustments in brightness and contrast. This term ensures the input image content is preserved and no mode-collapsing occurs.

Finally, (iii) we describe the use of an ensemble of relatively shallow discriminator networks in order to better match the low-level attributes of native LDR images, and suppress the edge-related artifacts that plague existing TMOs.
We demonstrate our method's ability to efficiently produce naturally-looking and artifact-free LDR renditions of highly challenging HDR scenes. A quantitative evaluation over benchmark HDR images reveals its superior performance over established image quality metrics and visual distances.



%% file: PrevWork.tex
\begin{figure*}[t]
\centering
   \includegraphics[width=0.9\linewidth]{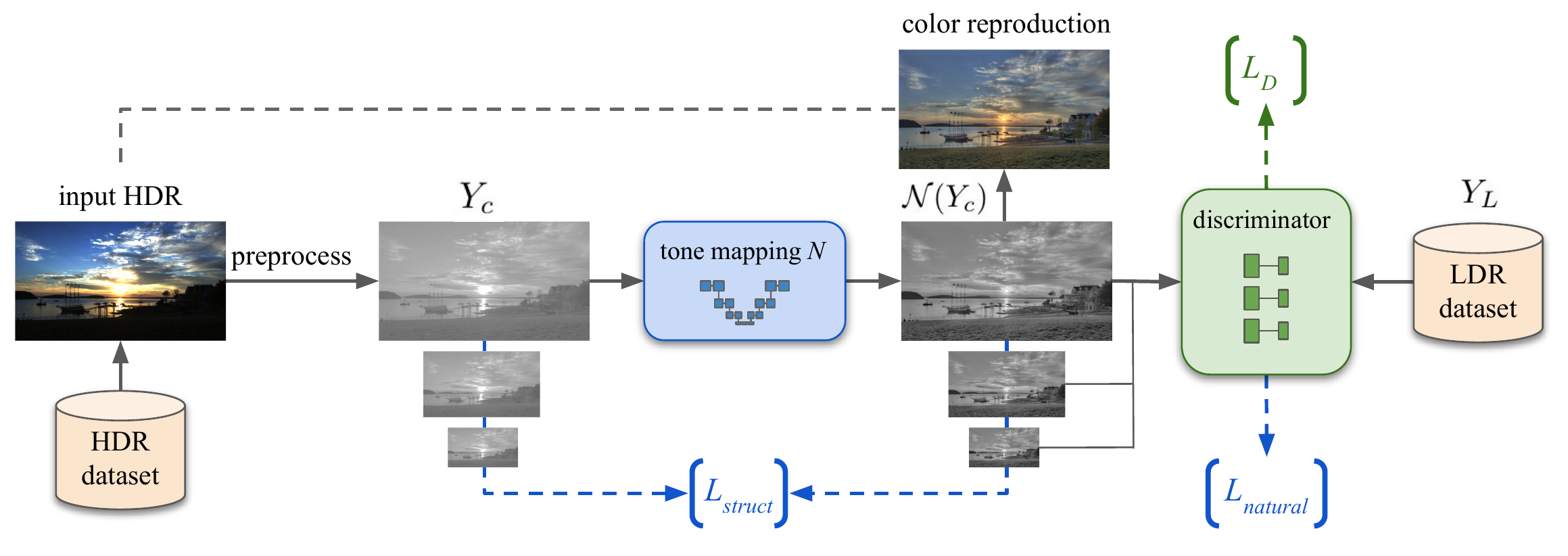}
  \caption{Tone-Mapping Pipeline. Given an input HDR image, we estimate the level of compression needed for mapping its luminance $\lum$ into a fixed-range map, $\loglum$. The latter is fed to the tone mapping network $\net$ as well as used for defining the structure-preservation loss. The network is also trained to minimize an adversarial loss with respect to native LDR images. The final output image is produced from the resulting luma, $\net(\loglum)$, and the chromaticity of the input image. We mark in blue and green the components that participate in the training of the network $\net$ and $\disc$ respectively.\vspace{-0.1in}}
  \label{fig:pipeline}
\end{figure*}

\section{Related Work}
\label{sec:prev}
We briefly review here the different tone-mapping approaches developed over the years, with an emphasis on recent DNN-based methods.

\bt{Tone-Mapping Algorithms} Tone mapping algorithms are typically divided into global and local operators. Global operators map pixels based on their tone and apply the same curve across the image. Notable examples use linear scaling~\cite{Ward94}, non-linear tone-curve~\cite{Tumblin93} and logarithmic tone-curve~\cite{Khademi2017,Drago2003}. The class of S-shaped curves is associated with responses in the human visual system (HVS)~\cite{Naka} and is used in~\cite{pattanaik2000time, Schlick94}. Computational efficiency and lack of over- and under-exposed pixels are the key advantages of global operators. However, their non-local reasoning leads to a sub-optimal local treatment characterized by a loss of contrast.
Local operators address this shortcoming by adapting their operation based on the local content of the image. Some works~\cite{Land_retinex,Pattanaik98,Reinhard02} derive their operation from the local adaptation mechanism of the HVS. However, the use of spatially-invariant filtering to analyze the image leads to unwanted halo effects around edges. 
To minimize these effects, different types of edge-aware filters were suggested, such as the anisotropic diffusion~\cite{Tumblin99}, robust averaging~\cite{DiCarlo00}, bilateral filter~\cite{Durand02}, weighted least-squares~\cite{Farbman08}, local-Laplacian pyramid~\cite{Paris11}, and a multiscale decomposition~\cite{Gu2013}. Another recent work by Liang~\etal~\cite{Liang18} uses a $l_0$ and $l_1$ sparsity to decompose image details. An additional approach tackles the problem by operating at the fine scale of image gradients~\cite{Fattal02, Shibata16}, and over a low-passed measure of contrast~\cite{Mantiuk06}.
This research effort led to a significant improvement in the ability to tone-map HDR images while preserving their local contrasts and minimizing edge-related artifacts. Nevertheless, as noted above, high compression levels may still result in artifacts or compromised realism due to contrast imbalances.

\bt{DNN-based Tone-Mapping} A number of approaches were suggested for utilizing DNNs for HDR tone-mapping, despite the lack of ground-truth training data. Hou \etal~\cite{Hou17} train a network to reproduce the input HDR image from its log-transformed luminance map, using a VGG-perceptual loss~\cite{Johnson16}. This method operates under the assumption that the compressive effect of the logarithm will persist under this norm and given the limited network capacity used. Their network must be trained specifically for each input image. Gharbi \etal~\cite{gharbi17} use a network architecture inspired by the bilateral grid to perform various image enhancement operations given input and output example pairs. The emphasis of this approach is on reproducing the enhancements in real-time.

Several recent methods~\cite{Cao_2020, montulet2019deep, TMO-Net, Patel2018, Rana20, Zhang19} train their tone-mapping network using conditional-GAN~\cite{pix2pix2017,wang2018pix2pixHD}, to obtain a more abstract metric of matching distributions. This conditional framework, however, requires pairs of input HDR and a corresponding tone-mapped LDR. These methods differ in the way they obtain their paired data and the additional similarity losses they use.
Patel \etal~\cite{Patel2018}, Rana \etal~\cite{Rana20} and Cao \etal~\cite{Cao_2020} construct their training pairs by applying a list of tone-mapping algorithms over each example HDR image and choosing the result with the highest tone mapped quality index (TMQI)~\cite{TMQI} as its ground-truth label. Panetta \etal~\cite{TMO-Net} train over a supervised low-light images dataset. These methods aid the convergence of their adversarial training using an additional reconstruction loss, Rana \etal and Cao \etal use the VGG-perception loss, Patel \etal use an L1 loss, and Panetta \etal use a combination of this loss with a gradient profile loss for better considering edge information.
Zhang \etal~\cite{Zhang19} incorporate a manual supervision by providing three photographers with a tone-mapping toolbox which they can use to obtain the most compelling result for each training image. The image with the highest TMQI score is then used as the training label. Rico \etal~\cite{montulet2019deep} use the MIT-Adobe 5K dataset~\cite{fivek} which was also produced by multiple (five) photographers that were allowed to apply different retouching operations to each training image. Rico \etal ended up using the tone-mapped results of only one of the experts as the training label, and they also employ a VGG-based reconstruction loss.

By contrast to these methods, we alleviate the need for paired training examples by training our network to reproduce the visual attributes of native LDR images.

Let us finally mention the line of work on low-light image enhancement methods. While these methods are not designed to cope with HDR images, they are also expected to manipulate (increase) the brightness of images. As these manipulations apply considerably weaker changes to the input image than HDR TMOs, global curves-based approaches are sufficient. Indeed, Guo and Li \etal ~\cite{Guo2020} predict an image-specific curve mapping using deep learning. Wang \etal ~\cite{Wang19} use retouched images to train a network to deal with underexposed images, and Jiang \etal ~\cite{Enlightengan_2021} train an unsupervised GAN for this purpose. Unlike our method, this work does not take any particular measure to allow it to apply significant levels of dynamic range compression.

%% file: Method.tex
\section{Method}
\label{sec:method}

We begin with a brief overview of our new tone-mapping pipeline, which is summarized in Figure~\ref{fig:pipeline}, and proceed with a detailed description of each of its steps.

Similarly to other TMOs~\cite{Fattal02,Schlick94,Tumblin99}, we perform the tone-mapping over the luminance channel $\lum$ of the input HDR in YUV color-space. 

We rely on the power of adversarial training to match the visual appearance of native LDR images. This type of training is however a delicate process, prone to various instabilities.
To ensure stable training, at the first step of our pipeline, we bring the input luminance $\lum$ into a fixed range by mapping it through a TRC that applies the proper amount of compression. This amount is estimated from the discrepancy between $\lum$'s histogram and a canonical LDR luminance distribution. 

The resulting image is then fed into the tone mapping network, $\net$, which is trained to reproduce the visual attributes of native LDRs and remove the biases created by the TRC applied. This is achieved by an adversarial training with respect to a dataset of native LDR images. We augment this training process with a loss that preserves the structural integrity of the input image by restricting the action of $\net$ to local changes in brightness and contrast. This loss also ensures no mode-collapsing occurs.

Finally, the output color image is recovered by the procedure used in~\cite{Schlick94,Tumblin99,Fattal02}.

\begin{figure*}
\centering
   \includegraphics[width=1.0\textwidth]{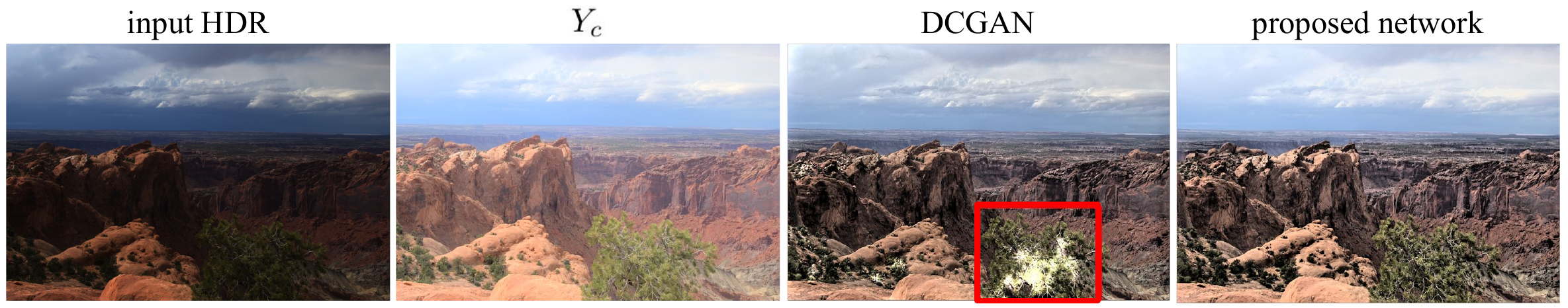}
  \caption{Operation of the Tone-Mapping Network. The figure shows an input HDR, a color rendition of $\loglum$ before being inputted to the network, and the action of $\net$ over $\loglum$. The latter reproduces a significant amount of contrasts which are missing in $\loglum$. The figure also compares the results obtained by two different discriminator networks. The deep DCGAN architecture leads to spurious brightened and darkened regions in the image, as indicated by the red frame. The use of an ensemble of shallower discriminator networks achieves a more consistent tone-mapping.\vspace{-0.15in}
  }
  \label{fig:dcgan}
\end{figure*}

\bt{Adaptive Curve-Based Compression} The luminance range of different HDR images can differ by orders of magnitude, which brings the need for applying different levels of compression. This variability in the input appears to challenge $\net$ and undermine the convergence of its training.
Consequently, as a pre-process step, we map the input luminances $\lum$ to a fixed range using a global curve mapping. The mapping ranges from a linear transformation to a severe log-like curve, depending on the level of compression required by the input $\lum$. This family of TRCs is given by
\begin{equation}
\label{eq:compress}
\loglum(\pix) = \log\left(\lumfac   \frac{\lum(\pix)}{ \max(\lum)}+\varepsilon\right)/ \log(\lumfac+\varepsilon),
\end{equation}
where $\max(\lum)$ refers to the maximal value across the image pixels $\pix$, the parameter $\lumfac$ sets the applied level of compression, and $\varepsilon=0.05$. The resulting luminance values are restricted to $[0,1]$ regardless of $\lumfac$. 

Logarithm curves are often used for tone-reproduction given their similarity to the compressive response to light that takes place in the human visual system, a.k.a. Weber-Fechner law~\cite{Drago2003}. We note however that unlike existing TRC-based methods we use an adaptive compression level, where different values of $\lumfac$ are used on different images. In Section~\ref{sec:luminance} below we describe the way we determine this parameter based on the histogram of $\lum$. 



\bt{Natural Appearance} Global TRCs, as the ones applied in Eq.~\ref{eq:compress}, are notorious for their inability to produce sufficient levels of local contrasts. To avoid this shortcoming as well as the question of what amount of contrasts should appear in the tone-mapped rendition, we adopt the goal of producing tone-mapped images that best resemble LDR images capturing native LDR scenes. 

We leverage the remarkable ability of adversarial training to reproduce a distribution given example images. More specifically, we train a discriminator network $\disc$ to distinguish between $\net(\loglum)$ and a set of high-fidelity native LDR images. By training $\net$ to fool $\disc$, its output acquires the desired appearance. The significant effect of the resulting $\net$ is demonstrated in Figure~\ref{fig:dcgan}.

This unpaired training scheme alleviates the need to collect or produce example tone-mapped images. As noted earlier, unlike~\cite{Patel2018,Rana20} who rely on existing tone mapping algorithms, our training allows us to surpass their performance. Being unsupervised, our training spares the manual effort assumed in~\cite{montulet2019deep,Zhang19}, as well as scores better on objective image fidelity measures.

The architecture of the discriminator network plays a central role in the success of this approach. The multi-layered DCGAN architecture prototype~\cite{radford15} has the ability to correlate between features at multiple scales and discriminate based on high-level semantic content. Consequently, this architecture is often used for training adversarial generative networks. We, however, do not train $\net$ to conceive new images from scratch, but specialize on removing biases in $\loglum$ with respect to regular LDR images.

Since these differences relate to local contrasts and edge-related effects, we set the focus of $\disc$ to this level of image modelling by limiting its depth to two convolutional layers. In order to identify discrepancies at multiple spatial scales, we follow the approach in~\cite{wang2018pix2pixHD}, and train an ensemble of discriminator networks, each one applied at a different image resolution, using the following loss
\begin{equation}
\begin{aligned}
\label{eq:disc_loss}
    \loss_{\disc} = \sum_{k\in\{0,1,2\}} \bigg( &\mathbb{E}_{\lumldr \sim LDR}\big[\disc_k(\downarrow^k\lumldr)-1\big]^2 \\ + &\mathbb{E}_{\lum \sim HDR}\Big[\disc_k\Big(\downarrow^k\net(\loglum)\Big)\Big]^2 \bigg),
\end{aligned}
\end{equation}
where $\lumldr$ denotes the luminance channel of the LDR training images, and $\loglum$ is the compressed luminance computed from $\lum $ using Eq.~\ref{eq:compress}. The $\downarrow^k$ denotes a bicubic $\times2^k$ image downscaling operator, and $\disc_k$ are the discriminator networks applied at the corresponding image scales. These networks have an identical architecture that we describe in Section~\ref{sec:arch} along with the architecture of $\net$.
Finally, we use $\disc_k$ to improve the ability of $\net$ to match the natural appearance of LDR images by training it to minimize
\begin{equation}
\begin{aligned}
\label{eq:natural}
    \loss_{natural} = \sum_{k\in\{0,1,2\}} \mathbb{E}_{\lum \sim HDR}\Big[\disc_k\Big(\downarrow^k\net(\loglum)\Big)-1\Big]^2,
\end{aligned}
\end{equation}

Note that both $\loss_{\disc}$ and $\loss_{natural}$ correspond to a least-squares GAN training losses~\cite{Mao17}. Figure~\ref{fig:pipeline} summarizes the computational graph related to these adversarial losses.

Finally, Figure~\ref{fig:dcgan} shows the stability related artifacts produced by the use of a potent deep DCGAN-styled discriminator compared to the more consistent results obtained by our ensemble of shallower networks.

\begin{figure}
\centering
   \includegraphics[width=0.45\textwidth]{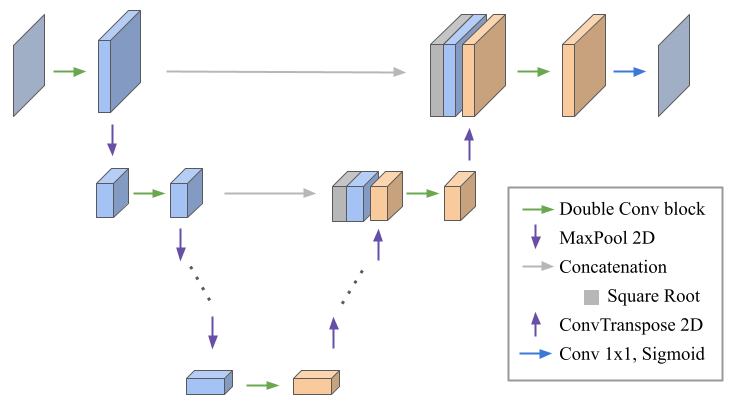}
  \caption{U-Net architecture that we use in our tone-mapping network $\net$. The skip connections at each layer pass the activations as well as their square-root as indicated by the gray boxes.\vspace{-0.15in}}
  \label{fig:unet}
\end{figure}


\bt{Structure Preservation} Mapping the luminance dynamic range into a narrower range inevitably changes the brightness of objects. This process can take place without changing the shape of the objects, in terms of the location and orientation of their edges.
However, the adversarial training in Eq.~\ref{eq:natural} does not require $\net(\loglum)$ to bear any similarity to the input $\lum$, or $\loglum$. To avoid this pitfall as well as any mode-collapsing during training, we promote the co-occurrence of brightness-normalized variations in the input $\loglum$ and the output $\net(\loglum) $ images.
We derive this measure based on the Pearson correlation inside small image patches of two images $I$ and $J$, by
\begin{equation}
\label{eq:struct_local}
\rho(I,J) = \frac{1}{n_p} \sum_{p_I,p_J} \frac{\mathrm{cov}(p_I, p_J)}{\sigma(p_I)\sigma(p_J)},
\end{equation}
where $p_I$ and $p_J$ are all the 5-by-5 pixel patches in $I$ and $J$ respectively. The covariance, $cov$, and standard deviation, $\sigma$, are also computed inside these patches. This measure is normalized by the total number of patches used, $n_p$. 

We use this measure to preserve the content and structural integrity of the input at multiple spatial scales by minimizing
\begin{equation}
\label{eq:struct}
\loss_{struct} = \sum_{k\in\{0,1,2\}} \rho(\downarrow^k\loglum,\downarrow^k\net(\loglum)),
\end{equation}
which we use in combination with the adversarial loss in Eq.~\ref{eq:natural} to train $\net$. 

We finally note that the structural similarity index (SSIM)~\cite{SSIM} also analyzes the images in the basis of its patches. However, it is not invariant to changes in the patch brightness and contrast. Hence, this index is irrelevant for our purpose as $\net(\loglum)$ is expected to undergo major changes in both brightness and contrast.

\bt{Color Reproduction} At inference time we recover an output color image using the formula used in~\cite{Fattal02,Schlick94,Tumblin99}. Specifically, each RGB color channel $C_{in}$ of the input HDR image is mapped independently to produce the corresponding output channel by, $C_{out}=(C_{in}/\lum)^s \net(\loglum)$. We use the default color saturation parameter $s=0.5$.


\subsection{Compression Level Estimation}
\label{sec:luminance}

The TRC we use in Eq.~\ref{eq:compress} allows us to apply different levels of compression, depending on the amount needed for each image. We explain here the way we determine this level.

When collecting training images from the wild, or applying our TMO over images from arbitrary sources, the meta-data related to the scale of their luminance values may not be available. Hence, determining $\lumfac$ based on the maximal $\lum$ value can be unreliable. The option of using the dynamic range, computed by the ratio between the largest and smallest input luminances, may be unstable as the noise level regulating this estimate is unavailable in such circumstances.

To avoid these indeterminacies and train, and apply, our method over arbitrary images, we derive the following estimate for $\lumfac$. Specifically, since $\lumfac$ controls the TRC in Eq.~\ref{eq:compress}, we search for the value that brings $\loglum$ closest to a canonical LDR image behaviour. As the luminance distribution of HDR images is considerably richer than the one of LDR images, we use these distributions as the descriptors for measuring this proximity. 

Formally, we search for the $\lumfac$ that minimizes the following cross-entropy
\begin{equation}
\label{eq:hist}
\min_{\lumfac} -\sum_{l} H_l(\loglum) \log \big(H_l(LDR)\big),
\end{equation}
where the histogram of $\loglum$, denoted by $H(\loglum)$, is a function of $\lumfac$, and $H(LDR)$ denotes the histogram of native LDR images. The latter was computed by averaging the histogram of 900 high-quality images from the DIV2k dataset~\cite{DIV2K}. We used 20 bins indexed by $l$. Finally, we solve the optimization in Eq.~\ref{eq:hist} using a stochastic search~\cite{Storn97}. 

Finally, in Section~\ref{sec:ablation} we report an ablation study demonstrating the contribution of this estimate compared to using the raw luminance values in the images we collected for training and testing our method.

\subsection{Implementation Details}
\label{sec:arch}

\bt{Tone-mapping Network Architecture} The network $\net$ consists of a U-Net architecture~\cite{Unet} with four levels, each containing two sets of 3-by-3 convolution, ReLU, and max-pooling. The number of filters is doubled at every level, starting with 32 filters. We use $\times2$ pooling factor and a convolution-transpose for unpooling. At the output layer, the ReLU operator is replaced by a sigmoid. This architecture is depicted in Figure~\ref{fig:unet}. 

The combination of linear layers (convolution) and ReLU operators span a space of piecewise affine transformations. In order to express a smoother luminance mapping, we concatenate the activations passing through the skip connections with their square-root, as indicated in Figure~\ref{fig:unet}. This step doubles the activations dimension, but the convolution that follows reduces this dimension. As shown in the ablation study in Section~\ref{sec:ablation}, these compressive pointwise transformations improve the photo-realism of our tone-mapped images.

\bt{Discriminator Architecture} We use three separate discriminators which do not share their weights. The networks have an identical architecture consisting of four layers. The first two are 4-by-4 convolution layers, with 16 and 32 filters respectively, and use LeakyReLU activations. Both levels have $\times2$ strided pooling. The third layer uses an unstrided 1-by-1 convolution, followed by an output fully-connected layer with a sigmoid activation.

\bt{Datasets} We train the network on HDR images taken from the HDR+ dataset~\cite{HDRp}, which contains indoor and outdoor scenes at different lighting conditions. We use 1000 images from this dataset for training and another 1000 images for testing. Each image is cropped and rescaled to provide two 256-by-256 images.
The native LDR images, used for the adversarial loss in Eq.~\ref{eq:disc_loss}, were taken from the DIV2k dataset~\cite{DIV2K}, which contains 1000 high-quality images. We randomly split this dataset into two halves for training and testing. Here again, we augment each set by cropping every image into two, and rescaling them to 256-by-256 pixels.

\bt{Training Details}
We implemented our model using the Pytorch framework and trained it on a single GeForce GTX1080Ti GPU using an Adam optimizer with a learning rate of $10^{-4}$ for $\net$ and $\times1.5$ this rate for $\disc$. We pre-trained $\disc$ to discriminate between $\loglum$ and LDR images for 50 epochs, and then trained both networks over 300 epochs with a learning-rate decay factor of half every 50 epochs. The pre-training stage is meant to stabilize the GAN training. The tone-mapping network $\net$ consists of about $4.5M$ parameters, and the discriminators a total of $31k$ parameters. Training these networks took approximately 3 and a half hours.
\bt{Running Times} It takes 0.5 seconds for our trained tone-mapping operator to process a 1333-by-2000 pixels image using a GPU. 




%% file: Results.tex
\section{Results}
\begin{figure*}
\centering
\includegraphics[width=1\textwidth]{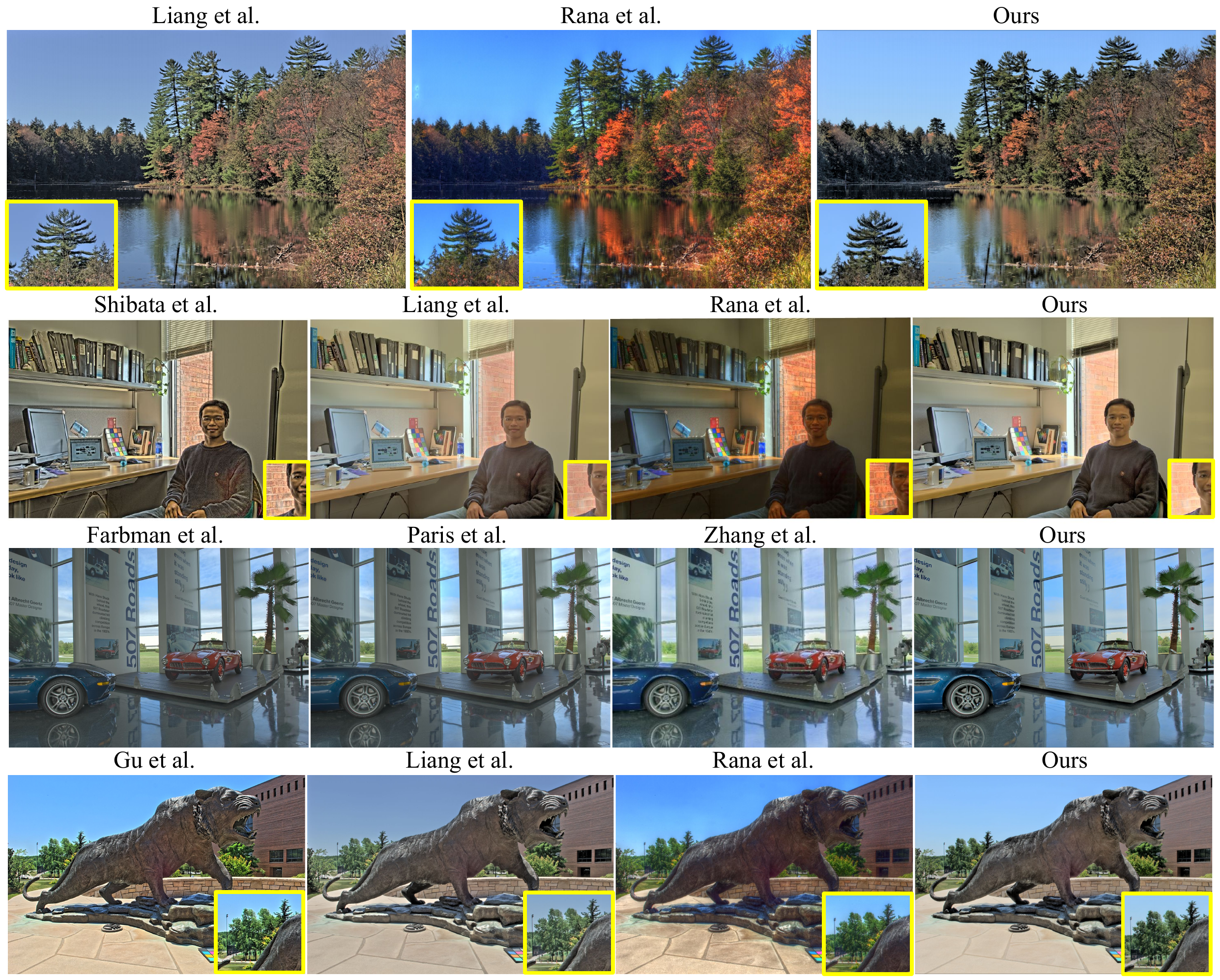}

  \caption{Tone-mapped images produced by different methods.\vspace{-0.15in}}
  \label{fig:comp}
\end{figure*}

We report the quantitative and qualitative evaluation of our tone-mapping network next to the state-of-the-art algorithms of Farbman 
 \etal~\cite{Farbman08}, Ferradans \etal~\cite{ferradans2011analysis}, Gu \etal~\cite{Gu2013}, Mai \etal~\cite{mai2010optimizing}, Shan \etal~\cite{shan2009globally}, Shibata \etal~\cite{Shibata16}, Paris \etal~\cite{Paris11}, Liang \etal~\cite{Liang18}, Khan \etal~\cite{Khan_2020} and Zhang \etal~\cite{Retina_inspired_2020}. The default parameters published by the authors were used for testing these methods.

We also compare to DNN-based methods of Zhang \etal~\cite{Zhang19}, Rana \etal~\cite{Rana20}, Panetta \etal~\cite{TMO-Net} and Cao \etal~\cite{Cao_2020}, by running our method on the same sets of test images these methods used in their evaluation.

 
\subsection{Quantitative Evaluation} 

In this comparison we used the tone-mapped image quality index (TMQI)~\cite{TMQI} and the blind TMQI (BTMQI)~\cite{BTMQI}, which are commonly used for assessing TMOs. These indices are evaluated over 105 benchmark HDR images from the HDR Photographic Survey dataset~\cite{HDRsurvey}. In cases where the competitors' original code was not available, we used their test images to evaluate our method using these scores.

The Fréchet Inception distance (FID)~\cite{FID} is a benchmark score for measuring the distance between distributions of visual descriptors extracted from the Inception network~\cite{inception}. We use this distance to measure the resemblance of our tone-mapped images and native LDR images. This distance, however, requires a large number of images ($\geq10,000$) which are currently not available in the domain of HDR imaging. Instead, we use a variant of this distance, which we call \emph{pixFID}, and can be reliably evaluated using less images. In this test we collected 1000 image from the HDR+ dataset~\cite{HDRp}, and another 1000 test LDR images from the DIV2k dataset~\cite{DIV2K}.
At the Appendix at the Supplemental Material we describe this distance and show its consistency with the FID.

\begin{table}[t]
\begin{center}
\begin{tabular}{lcc}
    \hline
    TMO & TMQI & BTMQI\\
    \hline
    Ferradans \etal~\cite{ferradans2011analysis} & 0.836 & 4.563\\
    Mai \etal~\cite{mai2010optimizing} & 0.856 & 3.958\\
    Shibata \etal~\cite{Shibata16} & 0.87 & 3.578 \\
    Gu \etal~\cite{Gu2013} & 0.871 & 3.878\\
    Shan \etal~\cite{shan2009globally} & 0.874 & 3.625 \\
    Zhang \etal~\cite{Retina_inspired_2020}* & 0.88 & 3.76 \\
    Rana \etal~\cite{Rana20}* & 0.88 & --\\
    Farbman \etal~\cite{Farbman08}& 0.886 & 3.602\\
    Liang \etal~\cite{Liang18} & 0.887 & 3.691 \\
    Khan \etal~\cite{Khan_2020}* & 0.889 & -- \\
    Ma \etal~\cite{TMQI_optim} & 0.895 & 3.868 \\
    Cao \etal~\cite{Cao_2020}* & 0.9 & -- \\
    Paris \etal~\cite{Paris11} & 0.906 & 2.988\\
    \textbf{Ours} & \textbf{0.919} & \textbf{2.89} \\
    \hdashline
    Zhang \etal~\cite{Zhang19}** & 0.874 & 3.519 \\
    Ours* & 0.902 & 3.24 \\
    \hdashline
    Panetta \etal~\cite{TMO-Net}*** & 0.873 & 3.44 \\
    Ours* & 0.883 & 3.041 \\
    \hline
\end{tabular}
\end{center}
\caption{Mean TMQI scores (higher is better) and BTMQI (lower is better), both computed over the same 105 images from the HDR Survey dataset~\cite{HDRsurvey}. In (*) these scores were taken directly from the respective paper. The scores in (**) were computed over 304 images from the HDRI Haven dataset~\cite{zaal}, which were used to evaluate~\cite{Zhang19}, and in (***) over the 456 images which were used for the evaluation in~\cite{TMO-Net}.\vspace{-0.15in}}
  \label{tab:TMQI}

\end{table}


\bt{Tone-Mapped Image Quality Indices} Our network is trained to reproduce the visual appearance of native LDR images while preserving the structural content of the input. Both the TMQI and BTMQI reward for these qualities and indeed, as Table~\ref{tab:TMQI} reports, our method achieves the best scores, with a non-negligible margin in the TMQI, next to a list of top-scoring tone-mapping algorithms and the recent DNN-based approaches of Zhang~\etal~\cite{Zhang19}, Rana~\etal~\cite{Rana20}, Panetta \etal~\cite{TMO-Net} and Cao \etal~\cite{Cao_2020}. Note that the latter DNN-based methods are not the next top scoring TMOs in the list.
Hence, beyond pushing the state-of-the-art in HDR tone-mapping, our approach does so using DNNs which, as we note below, are available in a more efficient platform.

\bt{Fréchet Inception Distance} Table~\ref{tab:FID} reports the pixFID values obtained by the different methods where our network receives the lowest (best) value. This suggests our tone-mapped images bear the closest statistical resemblance to native LDR images. Note that the recent algorithm of Liang \etal is our closest competitor in this score. This score was not evaluated over the DNN-based methods due to the unavailability of their code.

\subsection{Visual Evaluation} Figure~\ref{fig:comp} provides a side-by-side comparison of tone-mapped images produced by the methods achieving the top scores in the tests above.

The method of Shibata \etal appears to produce a proper compression level but also over-emphasized fine details and thin halos. The results of Liang \etal's method show less edge-related artifacts and an adequate compression, but contain a fairly modest amount of local and global contrasts. The method of Rana \etal produces halo-free images which are rich in contrast, but also contain overly dark regions due to an insufficient amount of compression. Both the algorithms of Farbman \etal and Paris \etal produce artifacts-free images, but seem to apply an insufficient level of compression that leads to weaker contrasts. Zhang \etal's results are better compressed and contain improved contrasts. Finally, the method of Gu \etal achieves a fair level of compression but suffers from  over-emphasized fine details and minor halo effects.
The tone-mapped images produced by our method do not seem to suffer from visual artifacts, and have an adequate amount of compression with no over- and under-exposed regions. The local and global contrasts appear fairly balanced, and consistent with ones found in contrast-rich native LDR images.
Additional side-by-side comparison images can be found in the Supplemental Material of this paper.

\textbf{User Study.} Finally, in order to further assess the visual quality of our results, we conducted a user-study that follows the protocol used in~\cite{Liang18}. Specifically, it consists of eight participants who were asked to rate between 1 to 8 images produced by the methods in~\cite{Liang18, Paris11, Rana20, Gu2013, Farbman08} as well as ours. The images were displayed next to each other (at a random order). The participants were naive as to the purpose of the test and were asked to score the images based on how realistic and unprocessed thy appear.

Our method obtained the highest mean score 6.43 with std 1.6, while the next top scoring method is \cite{Liang18} with mean score of 5.78 with standard-deviation of 2.04, and then \cite{Paris11} (4.39, 1.99), \cite{Farbman08} (4.70, 1.98), \cite{Gu2013} (2.43, 1.95) and \cite{Rana20} (2.14, 1.28). A plot summarizing these results can be found at the Supplemental Material.
\begin{table}[t]
\begin{center}
  \begin{tabular}{lc}
    \hline
    TMO & pixFID \\
    \hline
    Shan \etal~\cite{shan2009globally} & 1.22 \\
    Shibata \etal~\cite{Shibata16}& 1.20 \\
    Ferradans \etal~\cite{ferradans2011analysis} & 1.165 \\
    Mai \etal~\cite{mai2010optimizing} & 1.141 \\
    Paris \etal~\cite{Paris11} & 1.107 \\
    Gu \etal~\cite{Gu2013} & 1.105 \\
    Farbman \etal~\cite{Farbman08}& 1.102 \\
    Liang \etal~\cite{Liang18}& 1.04 \\
    \textbf{Ours} & \textbf{1.008} \\
    \hline
\end{tabular}
\end{center}
\caption{Pixel Fréchet Inception Distance (lower is better) computed against native LDR images from the DIV2k dataset.\vspace{-0.15in}}
  \label{tab:FID}
\end{table}

\subsection{Ablation Study} 
\label{sec:ablation}

In order to demonstrate the contribution of different components described in Section~\ref{sec:method}, we evaluated the TMQI and FID scores obtained by switching them off. For reference, we remind that our method achieves pixFID of 1.008 and TMQI of 0.919.

\textbf{Adaptive Compression Level.} TMOs that use a TRC determine the amount of compression in one of two ways: (i) operate on the raw luminance values of the input image, i.e., omitting the division by $\max(\lum)$ in Eq.~\ref{eq:compress}, and allow their range to affect the compression~\cite{Fattal02,Durand02,Farbman08}, or (ii) as in~\cite{Khademi2017} the input luminance is normalized to a fixed range, i.e., using a fixed amount of compression $\lumfac=5000$ in Eq.~\ref{eq:compress}.

In order to successfully cope with images of different dynamic range our method adapts the compression level it applies. Moreover, it estimates this level from the input histogram and avoids inconsistencies in the luminance scales of images curated from different sources. Using a fixed $\lumfac=1000$ lowered the TMQI score considerably to 0.89, and omitting the luminance normalization undermined both the pixFID to 1.096 and TMQI to 0.75.
We provide visual demonstrations at the Supplemental Material.

\textbf{Square-Root Transformations.} As described above, we augment our U-Net architecture with $\sqrt{x}$ transformations to better span smooth tone-mapping operators. By omitting these transformations the pixFID score increases to 1.219 and the TMQI decreases to 0.869. The visual impact of these transformations is substantial as we demonstrate in the Supplemental Material.

\textbf{Discriminator Ensemble.} We also evaluated the importance of using an ensemble of shallow discriminator networks applied at multiple image resolutions to two alternatives: (i) the use of a single shallow discriminator at the finest image resolution, and (ii) the use of a single deeper DCGAN architecture~\cite{radford15}. The first option results in inferior scores of pixFID 1.096 and TMQI 0.912, as well as a noticeable compromised ability to restore contrasts at coarse scales of the images as shown in the Supplemental Material. Figure~\ref{fig:dcgan} shows the artifacts resulting from the instabilities involved in training a deep DCGAN discriminator network, despite the fact that this option did not have a significant impact on the visual scores, namely, pixFID of 1.043 and TMQI of 0.92. 

\textbf{Adversarial Loss.} Without the GAN loss, the network is trained only to minimize the structural loss and results in an identity mapping that outputs the input tone-mapped images unchanged, the TMQI and pixFID scores under these settings were 0.8 and 1.131 correspondingly.

\textbf{Structural Preservation.} Finally, we report that the removal of our structural-preservation loss did not allow our training to converge.

%% file: Conclusions.tex
\section{Conclusions}
\label{sec:conc}

We presented a new training approach that, unlike exiting DNN-based methods, does not require paired training examples by setting the goal of producing naturally looking LDR renditions. In order to allow this minimal setting to succeed, we presented a range-normalizing pre-process which estimates and applies a different level of curve-based compression, a loss that restricts the action of the network to local changes in brightness and contrast, and a more concise discriminator network, designed to promote the reproduction of low-level attributes native LDR posses.
We evaluated our method on several image fidelity metrics and reported its superior performance. A visual inspection shows its ability to produce naturally-looking tone-mapping with balanced levels of contrast. 

%% file: egpaper_for_review.bbl
\begin{thebibliography}{10}\itemsep=-1pt

\bibitem{DIV2K}
Eirikur Agustsson and Radu Timofte.
\newblock Ntire 2017 challenge on single image super-resolution: Dataset and
  study.
\newblock In {\em The IEEE Conference on Computer Vision and Pattern
  Recognition (CVPR) Workshops}, July 2017.

\bibitem{fivek}
Vladimir Bychkovsky, Sylvain Paris, Eric Chan, and Fr{\'e}do Durand.
\newblock Learning photographic global tonal adjustment with a database of
  input / output image pairs.
\newblock In {\em The Twenty-Fourth IEEE Conference on Computer Vision and
  Pattern Recognition}, 2011.

\bibitem{Cao_2020}
X. Cao, K. Lai, S.N. Yanushkevich, and M.~R. Smith.
\newblock Adversarial and adaptive tone mapping operator for high dynamic range
  images.
\newblock In {\em 2020 IEEE Symposium Series on Computational Intelligence
  (SSCI)}, pages 1814--1821, 2020.

\bibitem{DiCarlo00}
Jeffrey DiCarlo and Brian Wandell.
\newblock Rendering high dynamic range images.
\newblock {\em Proceedings of SPIE - The International Society for Optical
  Engineering}, 3965:392--401, 05 2000.

\bibitem{Drago2003}
Fr{\'e}d{\'e}ric Drago, Karol Myszkowski, Thomas Annen, and Norishige Chiba.
\newblock Adaptive logarithmic mapping for displaying high contrast scenes.
\newblock {\em Comput. Graph. Forum}, 22:419--426, 2003.

\bibitem{Durand02}
Fr\'{e}do Durand and Julie Dorsey.
\newblock Fast bilateral filtering for the display of high-dynamic-range
  images.
\newblock In {\em Proceedings of the 29th Annual Conference on Computer
  Graphics and Interactive Techniques}, SIGGRAPH ’02, page 257–266, 2002.

\bibitem{HDRsurvey}
M.D. Fairchild.
\newblock {\em The HDR photographic survey}, pages 233--238.
\newblock 01 2007.

\bibitem{Farbman08}
Zeev Farbman, Raanan Fattal, Dani Lischinski, and Richard Szeliski.
\newblock Edge-preserving decompositions for multi-scale tone and detail
  manipulation.
\newblock {\em ACM Transactions on Graphics (Proceedings of ACM SIGGRAPH
  2008)}, 27(3):to appear, Aug. 2008.

\bibitem{Fattal02}
Raanan Fattal, Dani Lischinski, and Michael Werman.
\newblock Gradient domain high dynamic range compression.
\newblock {\em ACM Trans. Graph.}, 21:249–256, 2002.

\bibitem{ferradans2011analysis}
Sira Ferradans, Marcelo Bertalmio, Edoardo Provenzi, and Vincent Caselles.
\newblock An analysis of visual adaptation and contrast perception for tone
  mapping.
\newblock {\em IEEE Transactions on Pattern Analysis and Machine Intelligence},
  33(10):2002--2012, 2011.

\bibitem{gharbi17}
Micha{\"e}l Gharbi, Jiawen Chen, Jonathan~T Barron, Samuel~W Hasinoff, and
  Fr{\'e}do Durand.
\newblock Deep bilateral learning for real-time image enhancement.
\newblock {\em ACM Transactions on Graphics (TOG)}, 36(4):118, 2017.

\bibitem{Gu2013}
B. {Gu}, W. {Li}, M. {Zhu}, and M. {Wang}.
\newblock Local edge-preserving multiscale decomposition for high dynamic range
  image tone mapping.
\newblock {\em IEEE Transactions on Image Processing}, 22(1):70--79, 2013.

\bibitem{BTMQI}
Ke Gu, Shiqi Wang, Guangtao Zhai, Siwei Ma, Xiaokang Yang, Weisi Lin, Wenjun
  Zhang, and Wen Gao.
\newblock Blind quality assessment of tone-mapped images via analysis of
  information, naturalness, and structure.
\newblock {\em IEEE Transactions on Multimedia}, 18(3):432--443, Mar. 2016.

\bibitem{Guo2020}
Chunle Guo, Chongyi Li, Jichang Guo, Chen~Change Loy, Junhui Hou, Sam Kwong,
  and Runmin Cong.
\newblock Zero-reference deep curve estimation for low-light image enhancement.
\newblock In {\em Proceedings of the IEEE/CVF Conference on Computer Vision and
  Pattern Recognition (CVPR)}, June 2020.

\bibitem{HDRp}
Samuel~W. Hasinoff, Dillon Sharlet, Ryan Geiss, Andrew Adams, Jonathan~T.
  Barron, Florian Kainz, Jiawen Chen, and Marc Levoy.
\newblock Burst photography for high dynamic range and low-light imaging on
  mobile cameras.
\newblock {\em ACM Transactions on Graphics (Proc. SIGGRAPH Asia)}, 35(6),
  2016.

\bibitem{FID}
Martin Heusel, Hubert Ramsauer, Thomas Unterthiner, Bernhard Nessler, and Sepp
  Hochreiter.
\newblock Gans trained by a two time-scale update rule converge to a local nash
  equilibrium.
\newblock In I. Guyon, U.~V. Luxburg, S. Bengio, H. Wallach, R. Fergus, S.
  Vishwanathan, and R. Garnett, editors, {\em Advances in Neural Information
  Processing Systems 30}, pages 6626--6637. Curran Associates, Inc., 2017.

\bibitem{Hou17}
Xianxu Hou, Jiang Duan, and Guoping Qiu.
\newblock Deep feature consistent deep image transformations: Downscaling,
  decolorization and {HDR} tone mapping.
\newblock {\em CoRR}, abs/1707.09482, 2017.

\bibitem{pix2pix2017}
Phillip Isola, Jun-Yan Zhu, Tinghui Zhou, and Alexei~A Efros.
\newblock Image-to-image translation with conditional adversarial networks.
\newblock {\em CVPR}, 2017.

\bibitem{Enlightengan_2021}
Yifan Jiang, Xinyu Gong, Ding Liu, Yu Cheng, Chen Fang, Xiaohui Shen, Jianchao
  Yang, Pan Zhou, and Zhangyang Wang.
\newblock Enlightengan: Deep light enhancement without paired supervision.
\newblock {\em {IEEE} Trans. Image Process.}, 30:2340--2349, 2021.

\bibitem{Johnson16}
Justin Johnson, Alexandre Alahi, and Li Fei-Fei.
\newblock Perceptual losses for real-time style transfer and super-resolution.
\newblock In {\em European Conference on Computer Vision}, 2016.

\bibitem{Khademi2017}
Nima~Khademi Kalantari and Ravi Ramamoorthi.
\newblock Deep high dynamic range imaging of dynamic scenes.
\newblock {\em ACM Transactions on Graphics (Proceedings of SIGGRAPH 2017)},
  36(4), 2017.

\bibitem{Khan_2020}
Ishtiaq~Rasool Khan, Wajid Aziz, and Seong-O. Shim.
\newblock Tone-mapping using perceptual-quantizer and image histogram.
\newblock {\em IEEE Access}, 8:31350--31358, 2020.

\bibitem{Land_retinex}
Edwin Land and John McCann.
\newblock Lightness and retinex theory.
\newblock {\em Journal of the Optical Society of America}, 61:1--11, 02 1971.

\bibitem{Larson97}
Gregory~Ward Larson, Holly Rushmeier, and Christine Piatko.
\newblock A visibility matching tone reproduction operator for high dynamic
  range scenes.
\newblock {\em IEEE Transactions on Visualization and Computer Graphics}, pages
  291--306, 1997.

\bibitem{Liang18}
Zhetong Liang, Jun Xu, David Zhang, Zisheng Cao, and Lei Zhang.
\newblock A hybrid l1-l0 layer decomposition model for tone mapping.
\newblock In {\em The IEEE Conference on Computer Vision and Pattern
  Recognition (CVPR)}, 2018.

\bibitem{TMQI_optim}
Kede Ma, Hojatollah Yeganeh, Kai Zeng, and Zhou Wang.
\newblock High dynamic range image compression by optimizing tone mapped image
  quality index.
\newblock {\em IEEE Transactions on Image Processing}, 24(10):3086--3097, 2015.

\bibitem{mai2010optimizing}
Zicong Mai, Hassan Mansour, Rafal Mantiuk, Panos Nasiopoulos, Rabab Ward, and
  Wolfgang Heidrich.
\newblock Optimizing a tone curve for backward-compatible high dynamic range
  image and video compression.
\newblock {\em IEEE transactions on image processing}, 20(6):1558--1571, 2010.

\bibitem{Mantiuk06}
Rafal Mantiuk, Karol Myszkowski, and Hans-Peter Seidel.
\newblock A perceptual framework for contrast processing of high dynamic range
  images.
\newblock {\em ACM Trans. Appl. Percept.}, 3(3):286–308, 2006.

\bibitem{Mao17}
Xudong Mao, Qing Li, Haoran Xie, Raymond Lau, Wang Zhen, and Stephen Smolley.
\newblock Least squares generative adversarial networks.
\newblock pages 2813--2821, 10 2017.

\bibitem{montulet2019deep}
Rico Montulet and A. Briassouli.
\newblock Deep learning for robust end-to-end tone mapping.
\newblock In {\em BMVC}, 2019.

\bibitem{Naka}
KI Naka and WA Rushton.
\newblock S-potentials from colour units in the retina of fish (cyprinidae).
\newblock {\em The Journal of physiology}, 185(3):536—555, August 1966.

\bibitem{TMO-Net}
Karen Panetta, Landry Kezebou, Victor Oludare, Sos Agaian, and Zehua Xia.
\newblock Tmo-net: A parameter-free tone mapping operator using generative
  adversarial network, and performance benchmarking on large scale hdr dataset.
\newblock {\em IEEE Access}, 9:39500--39517, 2021.

\bibitem{Paris11}
Sylvain Paris, Samuel~W. Hasinoff, and Jan Kautz.
\newblock Local laplacian filters: Edge-aware image processing with a laplacian
  pyramid.
\newblock {\em ACM Trans. Graph.}, 30(4), 2011.

\bibitem{Patel2018}
Vaibhav Patel, Purvik Shah, and Shanmuganathan Raman.
\newblock A generative adversarial network for tone mapping hdr images.
\newblock {\em National Conference on Computer Vision, Pattern Recognition,
  Image Processing, and Graphics. NCVPRIPG}, pages 220--231, 2018.

\bibitem{Pattanaik98}
Sumanta~N. Pattanaik, James~A. Ferwerda, Mark~D. Fairchild, and Donald~P.
  Greenberg.
\newblock A multiscale model of adaptation and spatial vision for realistic
  image display.
\newblock SIGGRAPH ’98, page 287–298. Association for Computing Machinery,
  1998.

\bibitem{pattanaik2000time}
Sumanta~N Pattanaik, Jack Tumblin, Hector Yee, and Donald~P Greenberg.
\newblock Time-dependent visual adaptation for fast realistic image display.
\newblock In {\em Proceedings of the 27th annual conference on Computer
  graphics and interactive techniques}, pages 47--54, 2000.

\bibitem{radford15}
Alec Radford, Luke Metz, and Soumith Chintala.
\newblock Unsupervised representation learning with deep convolutional
  generative adversarial networks.
\newblock In Yoshua Bengio and Yann LeCun, editors, {\em 4th International
  Conference on Learning Representations, {ICLR} 2016, San Juan, Puerto Rico,
  May 2-4, 2016, Conference Track Proceedings}, 2016.

\bibitem{Rana20}
A. {Rana}, P. {Singh}, G. {Valenzise}, F. {Dufaux}, N. {Komodakis}, and A.
  {Smolic}.
\newblock Deep tone mapping operator for high dynamic range images.
\newblock {\em IEEE Transactions on Image Processing}, 29:1285--1298, 2020.

\bibitem{Reinhard02}
Erik Reinhard, Michael Stark, Peter Shirley, and James Ferwerda.
\newblock Photographic tone reproduction for digital images.
\newblock {\em ACM Trans. Graph.}, 21(3):267–276, 2002.

\bibitem{Unet}
O. Ronneberger, P.Fischer, and T. Brox.
\newblock U-net: Convolutional networks for biomedical image segmentation.
\newblock In {\em Medical Image Computing and Computer-Assisted Intervention
  (MICCAI)}, volume 9351 of {\em LNCS}, pages 234--241. Springer, 2015.
\newblock (available on arXiv:1505.04597 [cs.CV]).

\bibitem{Schlick94}
Christophe Schlick.
\newblock Quantization techniques for visualization of high dynamic range
  pictures.
\newblock pages 7--20. Springer-Verlag, 1994.

\bibitem{shan2009globally}
Qi Shan, Jiaya Jia, and Michael~S Brown.
\newblock Globally optimized linear windowed tone mapping.
\newblock {\em IEEE transactions on visualization and computer graphics},
  16(4):663--675, 2009.

\bibitem{Shen18}
Ziyi Shen, {Wei Sheng} Lai, Tingfa Xu, Jan Kautz, and {Ming Hsuan} Yang.
\newblock Deep semantic face deblurring.
\newblock In {\em Proceedings - 2018 IEEE/CVF Conference on Computer Vision and
  Pattern Recognition, CVPR 2018}, Proceedings of the IEEE Computer Society
  Conference on Computer Vision and Pattern Recognition, pages 8260--8269. IEEE
  Computer Society, Dec. 2018.

\bibitem{Shibata16}
Takashi Shibata, Tanaka Masayuki, and Masatoshi Okutomi.
\newblock Gradient-domain image reconstruction framework with intensity-range
  and base-structure constraints.
\newblock pages 2745--2753, 06 2016.

\bibitem{Storn97}
Rainer Storn and Kenneth Price.
\newblock Differential evolution – a simple and efficient heuristic for
  global optimization over continuous spaces.
\newblock {\em J. of Global Optimization}, 11(4):341–359, Dec. 1997.

\bibitem{inception}
Christian Szegedy, Wei Liu, Yangqing Jia, Pierre Sermanet, Scott Reed, Dragomir
  Anguelov, Dumitru Erhan, Vincent Vanhoucke, and Andrew Rabinovich.
\newblock Going deeper with convolutions.
\newblock In {\em Proceedings of the IEEE conference on computer vision and
  pattern recognition}, pages 1--9, 2015.

\bibitem{Tumblin93}
Jack Tumblin and Holly Rushmeier.
\newblock Tone reproduction for realistic images.
\newblock {\em Computer Graphics and Applications, IEEE}, 13:42--48, 1993.

\bibitem{Tumblin99}
Jack Tumblin and Greg Turk.
\newblock Lcis: A boundary hierarchy for detail-preserving contrast reduction.
\newblock SIGGRAPH ’99, page 83–90. ACM Press/Addison-Wesley Publishing
  Co., 1999.

\bibitem{Wang19}
Ruixing Wang, Qing Zhang, Chi-Wing Fu, Xiaoyong Shen, Wei-Shi Zheng, and Jiaya
  Jia.
\newblock Underexposed photo enhancement using deep illumination estimation.
\newblock In {\em Proceedings of the IEEE/CVF Conference on Computer Vision and
  Pattern Recognition (CVPR)}, June 2019.

\bibitem{wang2018pix2pixHD}
Ting-Chun Wang, Ming-Yu Liu, Jun-Yan Zhu, Andrew Tao, Jan Kautz, and Bryan
  Catanzaro.
\newblock High-resolution image synthesis and semantic manipulation with
  conditional gans.
\newblock In {\em Proceedings of the IEEE Conference on Computer Vision and
  Pattern Recognition}, 2018.

\bibitem{SSIM}
Zhou Wang, A.~C. Bovik, H.~R. Sheikh, and E.~P. Simoncelli.
\newblock Image quality assessment: From error visibility to structural
  similarity.
\newblock {\em Trans. Img. Proc.}, 13(4):600–612, Apr. 2004.

\bibitem{Ward94}
Greg Ward.
\newblock {\em A Contrast-Based Scalefactor for Luminance Display}, page
  415–421.
\newblock Academic Press Professional, Inc., USA, 1994.

\bibitem{TMQI}
H. {Yeganeh} and Z. {Wang}.
\newblock Objective quality assessment of tone-mapped images.
\newblock {\em IEEE Transactions on Image Processing}, 22(2):657--667, 2013.

\bibitem{zaal}
Greg Zaal.
\newblock Hdri haven dataset.
\newblock \url{https://hdrihaven.com/}.
\newblock [Online; accessed 14-November-2020].

\bibitem{Zhang19}
N. {Zhang}, C. {Wang}, Y. {Zhao}, and R. {Wang}.
\newblock Deep tone mapping network in hsv color space.
\newblock In {\em 2019 IEEE Visual Communications and Image Processing (VCIP)},
  pages 1--4, 2019.

\bibitem{Retina_inspired_2020}
Xian-Shi Zhang, Kai fu Yang, Jun Zhou, and Yong-Jie Li.
\newblock Retina inspired tone mapping method for high dynamic range images.
\newblock {\em Opt. Express}, 28(5):5953--5964, Mar 2020.

\end{thebibliography}
